\documentclass[fleqn,usenatbib]{mnras}

\usepackage[T1]{fontenc}

\usepackage{graphicx}	
\usepackage{amsmath}	
\usepackage{amssymb}	
\usepackage{blindtext}  
\usepackage{comment}
\usepackage{pdflscape}
\usepackage{multirow}
\usepackage{newtxtext,newtxmath}

\title[Merger locations of the Galactic BNSs]{The Galactic neutron star population II - Systemic velocities and merger locations of binary neutron stars}

\author[N. Gaspari et al.]{
Nicola Gaspari$^{1}$\thanks{E-mail: nicola.gaspari@live.it},
Andrew J. Levan$^{1,2}$,
Ashley A. Chrimes$^{3,1}$,
Gijs Nelemans$^{1,4,5}$
\\
$^{1}$Department of Astrophysics/IMAPP, Radboud University, PO Box 9010, 6500 GL Nijmegen, The Netherlands\\
$^{2}$Department of Physics, University of Warwick, Coventry CV4 7AL, UK\\
$^{3}$European Space Agency (ESA), European Space Research and Technology Centre (ESTEC), Keplerlaan 1, 2201 AZ Noordwijk, the Netherlands \\
$^{4}$SRON, Netherlands Institute for Space Research, Niels Bohrweg 4, 2333 CA Leiden, The Netherlands\\
$^{5}$Institute of Astronomy, KU Leuven, Celestijnenlaan 200D, 3001 Leuven, Belgium\\
}

\date{Accepted XXX. Received YYY; in original form ZZZ}

\pubyear{2023}

\begin{document}
\label{firstpage}
\pagerange{\pageref{firstpage}--\pageref{lastpage}}
\maketitle

\begin{abstract}
The merger locations of binary neutron stars (BNSs) encode their galactic kinematics and provide insights into their connection to short gamma-ray bursts (SGRBs). In this work, we use the sample of Galactic BNSs with measured proper motions to investigate their kinematics and predict their merger locations. Using a synthetic image of the Milky Way and its Galactic potential we analyse the BNS mergers as seen from an extragalactic viewpoint and compare them to the location of SGRBs on and around their host galaxies. We find that the Galactocentric transverse velocities of the BNSs are similar in magnitude and direction to those of their Local Standards of Rest, which implies that the present-day systemic velocities are not isotropically oriented and the peculiar velocities might be as low as those of BNS progenitors. Both systemic and peculiar velocities fit a lognormal distribution, with the peculiar velocities being as low as $\sim 22-157$ km s$^{-1}$. We also find that the observed BNS sample is not representative of the whole Galactic population, but rather of systems born around the Sun's location with small peculiar velocities. When comparing the predicted BNS merger locations to SGRBs, we find that they cover the same range of projected offsets, host-normalized offsets, and fractional light. Therefore, the spread in SGRB locations can be reproduced by mergers of BNSs born in the Galactic disk with small peculiar velocities, although the median offset match is likely a coincidence due to the biased BNS sample.
\end{abstract}




\section{Introduction}\label{sec:1}

The merger of a binary neutron star (BNS) can manifest itself with a variety of transient phenomena. This includes short-duration gamma-ray bursts (SGRBs) and their afterglows, gravitational waves, and kilonovae, as shown by the multi-messenger observations of GW\,170817 \citep{2017ApJ...848L..12A}. Although all of these transients can be used to inform the physics of the mergers \citep{2007PhR...442..166N,2007NJPh....9...17L,2017LRR....20....3M}, SGRBs occupy a privileged position due to their luminosity, which makes them the easiest to detect \citep{2012ApJ...746...48M,2020LRR....23....4B}. Consequently, SGRBs provide the largest sample for analysis, and to date we have more than three decades of literature that explores their connection to BNS mergers (\citealt{1989Natur.340..126E,1992ApJ...395L..83N}; for reviews see \citealt{2014ARAA..52...43B}). The evidence supporting BNS mergers as progenitors are both indirect, such as the lack of association with supernovae, the redshift distribution, the demographics and location of their host galaxies \citep{2014ARAA..52...43B}, as well as direct, namely the concurrent detection of GRB\,170817A and GW\,170817. It is important to note, however, that BNS mergers do not have a one-to-one relation with SGRBs, since it is likely that not all SGRBs are produced by BNS mergers \citep{1995MNRAS.275..255T,1998ApJ...494L..57Q,2006MNRAS.368L...1L,2008MNRAS.385.1455M,2008MNRAS.385L..10T,2020ApJ...895...58G}, and not all mergers produce a SGRB \citep{2022Natur.612..223R,2022PhRvD.105h3004S,2022AA...666A.174S}.

A key piece of evidence connecting BNS mergers to SGRBs is their location within the host galaxy. Upon their formation in core-collapse supernovae, neutron stars (NSs) receive natal kicks, as evidenced by the observed peculiar velocities of young Galactic pulsars \citep{2005MNRAS.360..974H,2017A&A...608A..57V}. When the NS is in a binary, the natal kick adds to the systemic recoil due to the mass loss \citep[also known as Blaauw kick,][]{1961BAN....15..265B,1961BAN....15..291B}, and results in a kick to the binary barycenter of up to several hundred km s$^{-1}$ \citep{2017ApJ...846..170T,2018MNRAS.481.4009V,2019MNRAS.486.3213A}. Combined with the gravitational-wave in-spiral time, which can be as long as several Gyr or more, BNS can therefore migrate and merge well outside their host galaxy of origin \citep{1998AA...332..173P,1998AA...332L..57B,1999ApJ...526..152F,1999MNRAS.305..763B,2002ApJ...570..252P,2003MNRAS.342.1169V,2006ApJ...648.1110B,2009ApJ...705L.186Z,2011MNRAS.413.2004C,2014ApJ...792..123B,2022MNRAS.514.2716M}. This is observed in the host-offset distribution of SGRBs, which occur at larger projected radii from their hosts than any other class of transient \citep{2010ApJ...708....9F,2013ApJ...776...18F,2020ApJ...904..190Z,2022ApJ...940...56F}, and it is not expected in other progenitor scenarios \citep[e.g. ][]{1999ApJ...526..152F,2011NewAR..55....1B,2014ApJ...792..123B}. Nevertheless, a modest fraction ($\sim$20 per cent) of SGRBs are apparently ‘hostless’ (namely there is no underlying galaxy nor a single galaxy to clearly assign as their host), and although this is not at odds with BNS mergers, it leaves open questions about the nature of largest offsets \citep{2010ApJ...722.1946B,2014MNRAS.437.1495T,2022MNRAS.515.4890O}. Possible explanations are that BNSs received high systemic kicks \citep{2022MNRAS.515.4890O}, received low systemic kicks \citep{2016MNRAS.456.4089B} but were born either in globular clusters \citep{2006NatPh...2..116G,2010ApJ...720..953L,2011MNRAS.413.2004C} or in the outer regions of the host \citep{2021MNRAS.503.5997P}, or that they simply reside in a faint and/or distant host which hasn't been correctly identified \citep[e.g.][]{2007MNRAS.378.1439L}.

Understanding the merger locations of BNSs is also important in the context of Galactic chemical enrichment. BNS mergers are thought to be important sites for the production of r-process elements (\citealt{1989Natur.340..126E,1999ApJ...525L.121F,1999A&A...341..499R,2017Natur.551...67P,2017Natur.551...80K}; for reviews see \citealt{2021RvMP...93a5002C}) due to the neutron-rich ejecta and kilonovae they produce. There have been efforts to understand Galactic r-process enrichment in the context of BNS mergers in the Milky Way \citep{1982ApL....22..143S,1989Natur.340..126E,1999ApJ...525L.121F,2004A&A...416..997A,2014MNRAS.438.2177M,2015MNRAS.447..140V,2015ApJ...807..115S,2015MNRAS.452.1970W,2016ApJ...829L..13B,2018IJMPD..2742005H,2018ApJ...855...99C,2019ApJ...875..106C,2023ApJ...943L..12K}, and r-process deposition on Earth has also been linked to kilonovae in the local few kpc \citep{2019ApJ...881L...4B,2021ApJ...923..219W}. Therefore, understanding the locations of BNS mergers has wide-ranging implications. 

In this paper, we combine two approaches to studying BNS mergers – the locations of SGRBs, and the Galactic BNS population. We evolve a sample of Galactic BNSs forwards in time through the Galactic potential to determine their future merger locations. We place the merger locations in the context of the Milky Way as seen externally, and compare these results with observations of SGRBs in and around their host galaxies. The paper is structured as follows. In Section~\ref{sec:2} we describe our Galactic BNS sample, model for the Galactic potential, and a prescription for producing a synthetic Milky Way image. Section~\ref{sec:3} analyses the present-day velocities of observed Galactic BNSs and discusses their implications for birth locations and velocities. Section~\ref{sec:4} presents results for BNS merger locations and their measurements as viewed from afar. Section~\ref{sec:5} outlines possible systematics in the methodology, before we summarise and conclude in Section~\ref{sec:6}. 

Throughout, magnitudes are reported in the AB system \citep{1982PASP...94..586O}, and a flat $\Lambda$CDM cosmology is adopted with $H_0=70$ km s$^{-1}$ Mpc$^{-1}$ and $\Omega_\text{m}=0.3$. 

\section{Models}\label{sec:2}

\subsection{Merger locations of the Galactic BNSs}\label{sec:2.1}

We collected our sample of Galactic BNSs from the ATNF catalogue \citep{2005AJ....129.1993M}\footnote{\url{http://www.atnf.csiro.au/research/pulsar/psrcat}}, and their properties are summarized in Table~\ref{tab:1} along with the estimated merger times $\tau_\text{gw}$ for a gravitational-radiation driven inspiral \citep{1964PhRv..136.1224P}. Out of the 15 confirmed BNSs (we exclude possible NS-WD binaries), 8 have measured proper motions, and 5 have both proper motions and $\tau_\text{gw}<14\text{ Gyr}$. Since our primary objective is to make predictions about BNSs merging within an Hubble time, only the last 5 are employed in our fiducial models. The remaining BNSs are used to test for systematic effects in our methodology. 

To predict the BNSs merger locations we produce $10^4$ realizations of the Galactic trajectory of each binary, starting from as many realizations of their present-day positions and velocities. These initial conditions are generated through a Monte Carlo (MC) simulation, which employs observational uncertainties and allows us to propagate them to the predicted merger locations.

\subsubsection{Initial positions and velocities}\label{sec:2.1.1}

\begin{table*}
    \centering
	\caption{BNS sample used in this work. The columns list the pulsar, the right ascension and declination, the proper motion in right ascension and declination, the distance from the Sun, the merger time, and the references for the listed values. Values in parenthesis are the uncertainties in the preceding digits.}
	\label{tab:1}
	\begin{tabular}{lccccccc}
		\hline\hline
		& R.A. & Dec. & $\mu_\alpha$ & $\mu_\delta$ & Dist. & $\tau_\text{gw}$ & Ref. \\
		Radio pulsar & [deg] & [deg] & [mas yr$^{-1}$] & [mas yr$^{-1}$] & [kpc] & [Myr] & \\
		\hline
        J0737-3039A/B   & 114.46353508(11)  & -30.66130953(3)  & -3.82(62) & 2.13(23)    & 1.1(1,2) & 85 & (a)     \\
        B1534+12        & 234.291507208(13) & 11.932064964(17) & 1.482(7)  & -25.285(12) & 1.051(5)& 2735 & (b)     \\
        J1756-2251      & 269.19430755(7)   & -22.866486(6)    & -2.42(8)  & 0(20)       & 0.73(24,60)& 1457 & (c)      \\
        B1913+16        & 288.86666425(13)  & 16.10760744(14)  & -0.72(11) & -0.03(14)   & 4.1(7,20)& 301 & (d)     \\
        B2127+11C$^\star$       & 322.5050175(5)    & 12.1772803(12)   & -1.3(5)   & -3.3(10)    & 25.000& 215 & (e)     \\
        J0453+1559      & 73.4392237(3)     & 15.9892517(17)   & -5.5(5)   & -6.0(42)    & 0.522$^\dagger$ -- 1.07$^\ddagger$ & 1456721 & (f)     \\
        J1411+2551      & 212.828608(13)    & 25.852331(20)    & -3(12)    & -4(9)       & 1.131$^\dagger$ -- 0.98$^\ddagger$ & $>$465446 & (g)     \\
        J1518+4904      & 229.56999618(7)   & 49.07618089(5)   & -0.67(4)  & -8.53(4)    & 0.964$^\dagger$ -- 0.63$^\ddagger$ & $>$8826539 & (h)     \\
        \hline
        J0514-4002A$^{\star\S}$    & 78.5278863(9)     & -40.0469147(6)   & 5.19(22)  & -0.56(25)   & 25.000& 507733 & (i)     \\
        J0509+3801      & 77.3824504(11)    & 38.021690(4)     & -         & -           & 1.562$^\dagger$ & 579 & (j)     \\
        J1757-1854      & 269.2657683(3)    & -18.9009378(20)  & -         & -           & 19.559$^\dagger$ & 76 & (k)     \\
        J1811-1736      & 272.979308(13)    & -17.61047(12)    & -         & -           & 4.419$^\dagger$ & $>$1794804 & (l)     \\
        J1829+2456      & 277.3944450(9)    & 24.9383869(9)    & -         & -           & 0.909$^\dagger$ & $>$55375 & (m)     \\
        J1913+1102      & 288.3710592(13)   & 11.034928(3)     & -         & -           & 7.140$^\dagger$ & $>$465 & (n)     \\
        J1930-1852      & 292.623815(3)     & -18.862853(17)   & -         & -           & 2.004$^\dagger$ & $>$1e8 & (o)     \\
        J1946+2052      & 296.55888(3)      & 20.87351(3)      & -         & -           & 3.510$^\dagger$& $>$46 & (p)     \\
        J1753-2240$^\S$    & 268.41603(3)      & -22.6783(3)      & -         & -           & 3.232$^\dagger$ & - & (q)     \\
        J1755-2550$^\S$    & 268.910000(17)    & -25.8394(5)      & -         & -           & 4.891$^\dagger$ & - & (r)     \\
        J1759+5036$^\S$    & 269.940300(13)    & 50.615822(6)     & -         & -           & 0.543$^\dagger$ & $>$179604 & (s)     \\
        J1807-2459B$^{\star\S}$    & 271.8369634(3)    & -25.000532(5)    & -         & -           & 3.045$^\dagger$ & 1039373 & (t)     \\
        J1906+0746$^\S$    & 286.70358(17)     & 7.77386(20)      & -         & -           & 7.4(14,25) & 309 & (u)      \\
		\hline
	\end{tabular}
	\begin{flushleft}
	   \textbf{Notes.} $^\star$ Associated to a GC. $^\S$ Not a confirmed BNS. $^\dagger$ Estimated from DM following \cite{2017ApJ...835...29Y}. $^\ddagger$ Estimated from DM following \cite{2002astro.ph..7156C}.
	    
	   \textbf{References.} 
	   (a) \cite{2003Natur.426..531B, 2006Sci...314...97K, 2009Sci...323.1327D, 2012ApJ...755...39V}.
	   (b) \cite{2002ApJ...581..501S, 2014ApJ...787...82F}.
	   (c) \cite{2014MNRAS.443.2183F}.
	   (d) \cite{2010ApJ...722.1030W, 2016ApJ...829...55W, 2018ApJ...862..139D}.
	   (e) \cite{2006ApJ...644L.113J}.
	   (f) \cite{2015ApJ...812..143M}.
	   (g) \cite{2017ApJ...851L..29M}.
	   (h) \cite{2008AA...490..753J}.
	   (i) \cite{2019MNRAS.490.3860R}.
	   (j) \cite{2018ApJ...859...93L}.
	   (k) \cite{2018MNRAS.475L..57C}.
	   (l) \cite{2000AA...358L..53M, 2007AA...462..703C}.
	   (m) \cite{2004MNRAS.350L..61C, 2005MNRAS.363..929C}
	   (n) \cite{2016ApJ...831..150L}.
	   (o) \cite{2015ApJ...805..156S}.
	   (p) \cite{2018ApJ...854L..22S}.
	   (q) \cite{2009MNRAS.393..623K}.
	   (r) \cite{2018MNRAS.476.4315N}.
	   (s) \cite{2021ApJ...922...35A}.
	   (t) \cite{2012ApJ...745..109L}.
	   (u) \cite{2015ApJ...798..118V}.
	\end{flushleft}
\end{table*}
\begin{table*}
    \centering
	\caption{Properties of the GCs associated with a Galactic BNS. The columns list the pulsar, the associated GC, the right ascension and declination, the distance from the Sun, the proper motion in right ascension and declination, and the mean radial velocity. Values in parenthesis are the uncertainties in the preceding digits. Data are taken from \citet{2019MNRAS.482.5138B}.}
	\label{tab:2}
	\begin{tabular}{lccccccc}
		\hline\hline
		& & R.A. & Dec. & Dist. & $\mu_\alpha$ & $\mu_\delta$ & $V_\text{r}$ \\
		Radio pulsar & GC & [deg] & [deg] & [kpc] & [mas yr$^{-1}$] & [mas yr$^{-1}$] & [km s$^{-1}$] \\
		\hline
        B2127+11C       & NGC 7078  & 322.493042    & 12.167001     & 10.22(13) & -0.63(1)  & -3.80(1)  & -106.76(25) \\
        \hline
        J0514-4002A $^\S$ & NGC 1851  & 78.528160     & -40.046555    & 11.33(19) & 2.12(1)   & -0.63(1)  & 320.30(25) \\
        J1807-2459B$^\S$ & NGC 6544  & 271.835750    & -24.997333    & 2.60(27)  & -2.34(4)  & -18.66(4) & -38.12(76) \\
		\hline
	\end{tabular}
	\begin{flushleft}
        \textbf{Notes.} $^\S$ Not a confirmed BNS.
    \end{flushleft}
\end{table*}

\begin{figure}
	\centering
	\includegraphics[scale=1]{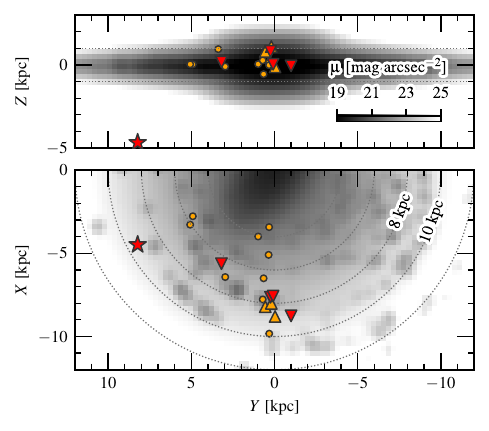}
    \caption{Present-day positions of the Galactic BNSs over a synthetic Milky Way image. Triangles indicate binaries with measured proper motions, while red markers indicate binaries merging within a Hubble time. Dots indicates binaries without measured proper motions. The star-shaped marker indicates B2127+11C, which is associated with the globular cluster NGC 7078. The underlying image is our fiducial model for a Milky Way image if it were located at $z=0.5$ and observed with a PSF FWHM of 0.1 arcsec, and a pixel size of 0.05 arcsec px$^{-1}$ (see Section \ref{sec:2.2} for details). The centre of the edge-on image reaches 18.3 mag arcsec$^{-2}$, and both images are cut at a limiting surface brightness of 25 mag arcsec$^{-2}$.
    }
    \label{fig:1}
\end{figure}

To compute the BNSs initial positions, we assume that right ascension, declination, and distance follow Gaussian distributions with mean equal to the estimated values and standard deviation equal to the respective uncertainties. These distributions are sampled $10^4$ times for each binary, giving as many realizations of their initial position. 

The distances listed in Table~\ref{tab:1} without uncertainties are estimates obtained from the dispersion measure (DM) using the electron-density model of \citet{2017ApJ...835...29Y}. For J0453+1559, J1411+2551, and J1518+4904, we also report the DM distances obtained with the model of \cite{2002astro.ph..7156C}, which will be used in Sec. \ref{sec:5.0} for comparison. On the DM distances we assume a conservative $20$ per cent uncertainty, as done by \citet{2017ApJ...846..170T}. For B2127+11C, as it is likely bound to the globular cluster (GC) NGC 7078 \citep{2014AA...565A..43K}, we assume its position to be that of the GC (see Table~\ref{tab:2}). Another exception is made for the distances of J0737-3039A/B and J1756-2251, for which we use the non-Gaussian probability distributions given by \citet{2012ApJ...755...39V}\footnote{\url{http://psrpop.phys.wvu.edu/LKbias/}}. The BNS present-day positions are shown in Galactocentric coordinates in Fig~\ref{fig:1}.

To compute the BNS initial velocities, we apply the same MC approach to the proper motions in right ascension and declination. We assume that the proper motions have Gaussian uncertainties with standard deviation equal to the observational uncertainties, and we produce $10^4$ realizations for each binary. Each realization is then converted to linear units (i.e. [km s$^{-1}$]) using one of the distance realizations, and the transverse component of the Sun's velocity is added to obtain the BNS transverse velocity $V_\text{t}$ in the Galactocentric frame. Since we have no observational estimates for the radial velocities $V_\text{r}$ but for B2127+11C (for which we use the radial velocity of NGC 7078), we obtain the 3D velocities through a MC simulation of their orientations $\theta$ with respect to the line-of-sight (LoS). 

For our fiducial models, we assume that the BNSs systemic velocities are isotropically oriented in the Galactocentric frame, and we hence compute $V_\text{r}$ as
\begin{equation}
    \label{eq:1}
    V_\text{r}=V_\text{t}\cot\theta
\end{equation}
where $\theta=\arccos{u}$ and $u$ is a real value uniformly sampled $10^4$ times between $0$ and $1$. Hereafter, we will refer to the systemic velocities obtained with this assumption as Galactocentric-isotropic velocities. If the sample were bigger, we could test the isotropy assumption e.g. by comparing the mean value of 1D velocities to that of 2D velocities, as done for isolated pulsars by \cite{2005MNRAS.360..974H}. However, the small sample size prevents us from doing so, hence the sole purpose of this assumption is to best reflect our ignorance about the radial velocities. 

\begin{figure*}
	\centering
    \includegraphics[scale=1]{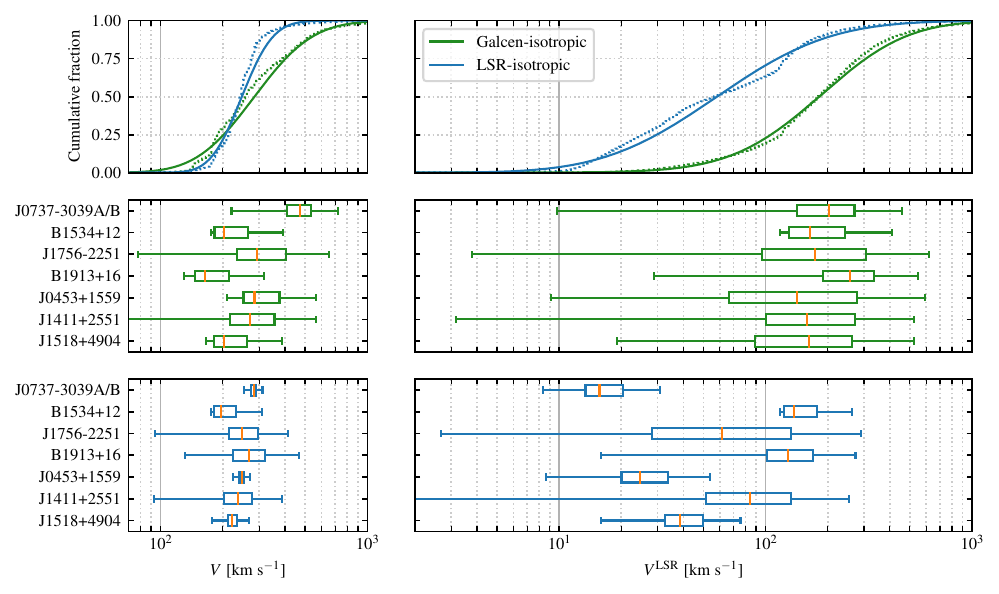}
    \caption{Galactocentric systemic velocities $V$ and peculiar velocities in the BNS LSRs $V^\textrm{LSR}$ of the Galactic BNSs with measured proper motions (except B2127+11C). The upper panels show the total cumulative distributions of $V$ and $V^\textrm{LSR}$. Dashed lines are distributions from the MC simulations, while solid lines are lognormal distribution fitted to the MC simulations. The middle and bottom panels show respectively the Galactocentric-isotropic and LSR-isotropic velocities for each single BNS considered. The boxes extend from the 1st to the 3rd quartiles, with the orange line on the median value.
    }
    \label{fig:2}
\end{figure*}

We also provide a second estimate for the radial velocities, obtained assuming that the peculiar velocities in the BNSs Local Standards of Rest (LSRs) are isotropically oriented. Here we define as BNS LSR the frame of reference centered on the BNS location, and moving on a circular orbit around the $Z$-axis of the Galactocentric frame. Under this assumption, we get $V_\text{r}$ by first subtracting the LSR transverse velocity vector ${\bf V}_\text{t,\,LSR}$ from the BNS transverse velocity vector ${\bf V}_\text{t}$, and then computing $V_\text{r}$ from the residuals, namely
\begin{equation}
    \label{eq:2}
    V_\text{r}=\|{\bf V}_\text{t}-{\bf V}_\text{t,\,LSR}\|\cot\theta+V_\text{r,\,LSR}
\end{equation} 
where $V_\text{r,\,LSR}$ is the BNS LSR radial velocity. Hereafter, we refer to these realizations as LSR-isotropic velocities. This second estimate is motivated by the possibility that Galactic BNSs might receive small systemic kicks from the second supernova \citep{2016MNRAS.456.4089B}, which would result in small peculiar velocities \citep{2019MNRAS.486.3213A}. The systemic velocities $V$ and the peculiar velocities $V^\text{LSR}=\|{\bf V}-{\bf V}_\text{LSR}\|$ obtained under the two assumptions are shown in Fig.~\ref{fig:2}.

The initial conditions are computed using the default values for the Sun Galactocentric position and velocity from \texttt{astropy v4.0}\footnote{\url{http://www.astropy.org}} \citep{astropy:2022}.

\begin{table}
    \centering
    \caption{Parameters of the lognormal distributions fitted to systemic velocities $V$ and peculiar velocities $V^\textrm{LSR}$. Last four rows reports median values $\mu$ and 16th--84th percentiles, using DM distances from either \citet{2017ApJ...835...29Y} (two upper rows) or \citet{2002astro.ph..7156C} (two lower rows) for J0453+1559, J1411+2551, and J1518+4904.
    }
    \label{tab:3}
    \begin{tabular}{lccccc}
    \hline\hline
    & \multicolumn{2}{c}{$V$} && \multicolumn{2}{c}{$V^\textrm{LSR}$} \\
    \cline{2-3}\cline{5-6}
    \rule{0pt}{3ex} & Galcen-iso & LSR-iso && Galcen-iso & LSR-iso \\
    \hline
    $\alpha$ & 8.48 & 21.35 && -11.14 & 0.46 \\
    $\beta$ & 276.16 & 226.13 && 192.91 & 57.66 \\
    $\gamma$ & 0.53 & 0.30 && 0.74 & 1.01 \\
    $\mu$ [km s$^{-1}$] & 285 & 245 && 182 & 58 \\
    $1\sigma$ [km s$^{-1}$] & 172-475 & 189-326 && 82-391 & 22-157 \\
    \hline
    $\mu$ [km s$^{-1}$] & 284 & 247 && 182 & 60 \\
    $1\sigma$ [km s$^{-1}$] & 175-474 & 192-323 && 83-391 & 23-159 \\
    \hline
    \end{tabular}
\end{table}

\subsubsection{Galactic trajectories}\label{sec:2.1.2}

The Galactic trajectories defined by each realization of initial position and velocity are computed with \texttt{galpy}\footnote{\url{https://github.com/jobovy/galpy}} \citep{2015ApJS..216...29B} using the Galactic potential model of \citet{2017MNRAS.465...76M}. The trajectories are evolved up to the merger time $\tau_\textrm{gw}$ of the respective binary, and the final positions are assumed to be the merger location. As we start from $10^4$ initial conditions for each BNS, we end up with the same number of merger locations. 

\subsection{Synthetic image of the Milky Way}\label{sec:2.2}

To analyze the BNS merger locations in the same way as SGRBs on their host, we need to reproduce how the Milky Way appears from a cosmological distance. To do so, we employ the Milky Way synthetic image of \citet{2021MNRAS.508.1929C}, upgrading their 2D face-on model to 3D so that we can include the effects of a random viewing angle.

\subsubsection{Model structure: bulge, disc, and spiral arms}\label{sec:2.2.1}

In the following, we provide a brief overview of the Galactic components combined to create the synthetic image, while a thorough description can be found in \citet{2021MNRAS.508.1929C}.

The bar-bulge is modelled with the triaxial boxy Gaussian distribution fitted on Mira variables
by \cite{2020MNRAS.492.3128G}, i.e. 
\begin{equation}
    \rho_\text{bar} = \rho_\text{b,0}\:\exp (-0.5 m^2)
\end{equation}
where
\begin{equation}
    m = \left\{\left[ \left( \frac{x}{X_\text{b}} \right)^2 + \left( \frac{y}{Y_\text{b}} \right)^2 \right]^2 + \left( \frac{z}{Z_\text{b}} \right)^4\right\}^{\frac{1}{4}}
\end{equation}
with $(X_\text{b},Y_\text{b},Z_\text{b})=(2.05,0.95,0.73)$ kpc, and $\rho_\text{b,0}$ is the normalization factor. The angle between the bar-bulge semi-major axis and the Galactic-centre LoS is assumed to be $27\degr$ \citep{2013MNRAS.435.1874W}. 

The disc is modelled with a double exponential disc 
\begin{equation}
\label{eq:5}
    \rho_\text{disc} = \rho_\text{d,0}\:\exp \left( -\frac{\sqrt{x^2+y^2}}{R_\text{d}} \right) \:\exp \left( -\frac{|z|}{Z_\text{d}} \right)
\end{equation}
where $R_\text{d}=2.6$ kpc and $Z_\text{d}=0.3$ kpc. These scale values are typical estimates for the Milky Way thin disc, which is the dominant component of the disc stellar light \citep{2016ARAA..54..529B}. 

The spiral arms are mapped following the method of \cite{2019ApJ...885..131R}, using young stellar object masers and \ion{H}{ii} regions from \cite{2014MNRAS.437.1791U} as tracers. To avoid the shadow produced by dust absorption behind the bar-bulge, we only use the lower half of the map, at Galactocentric coordinates $X\leq0$ (see Fig.~\ref{fig:1}), while the upper half is replaced by a reflected lower half.

The synthetic images are produced in two bands, namely the \textit{I}- and the \textit{B}-band. The normalization factors $\rho_\text{b,0}$ and $\rho_\text{d,0}$ are tuned so that the total luminosity of the respective components matches the observational estimates. In the \textit{I}-band we match the luminosities given by \cite{2006MNRAS.372.1149F}, namely $10^{10}\text{ L}_{\sun}$ for the bar-bulge, and $3\-\times\-10^{10}\text{ L}_{\sun}$ for the disc including spiral arms. For the \textit{B}-band we use instead the $B-I$ colours of Milky Way analogues from \cite{2015ApJ...809...96L}. 
We use $B-I=2.41$ for the bulge-bar and $B-I=1.62$ for the disc including arms, which give respectively a total luminosity of $0.11\-\times\-10^{3}\text{ L}_{\sun}$ and $0.67\-\times\-10^{3}\text{ L}_{\sun}$. The \textit{I}-band luminosities are corrected for dust extinction, while the \textit{B}-band luminosities are not \citep[for details see][]{2021MNRAS.508.1929C}.

The fraction of disc luminosity arising from the spiral arms alone is assumed to be equal to the arm strength, which is the relative amplitude of the 2nd to 4th order Fourier components of the azimuthal light profile along elliptical isophotes \citep[e.g.][]{2018ApJ...862...13Y}. 
For the \textit{I}-band we use an arm strength of $0.15$ \citep[e.g.][]{2019AA...631A..94D,2020ApJ...900..150Y}, while for the \textit{B}-band we use $0.20$ \citep{2018ApJ...862...13Y}, both of which are typical values for Milky Way analogues.

\subsubsection{Photometry and half-light radius}\label{sec:2.2.2}

The method described in the previous Section gives an analytical model of the Milky Way luminosity density in units of e.g. $[\text{L}_{\sun}\text{ pc}^{-3}]$. To produce a 2D image, the model is first projected along an arbitrary LoS, then processed to mimic the effects of redshift and instrumental resolution to simulate observations with a given instrument of the galaxy as viewed at an arbitrary redshift.

We start with a grid of points in Galactocentric coordinates. The grid is rotated by a phase $\phi$ along the $Z$-axis and by a inclination $i$ along the $Y$-axis, in order to be aligned with the chosen LoS. The analytical 3D model is evaluated on the grid and summed along the $X$-axis with a Riemann sum, to get a 2D image of the surface brightness $I$ in units of $[\text{L}_{\sun}\text{ pc}^{-2}]$. The number of grid points is chosen such that a double Riemann sum of the image gives a total luminosity differing by less than $1$ per cent from the prescribed value.

We then chose the observer redshift $z$, the point spread function (PSF), and the pixel scale of the image (i.e. the angular resolution). The 2D grid is converted from linear to angular units (e.g. from $[\text{pc}]$ to $[\text{arcsec}]$) using the angular diameter distance $D_\text{A}(z)$, and the image is first convolved with a Gaussian PSF of given full width at half maximum (FWHM) and then downsampled to the pixel scale. The last step is done using \texttt{measure.block\_reduce} from the \texttt{scikit-image} library \citep{scikit-image}. For the fiducial models we use $z=0.5$, a pixel size of $0.05\text{ arcsec}\text{ px}^{-1}$, and a PSF FWHM of $0.1\text{ arcsec}$. The fiducial redshift is chosen because the median redshift of our SGRB sample is $z=0.46$, while the other two values are typical for the SGRB hosts observations, for which the comparison sample used here is predominantly taken via the \textit{Hubble Space Telescope} (\textit{HST}, see Table~\ref{tab:4} for references). The possible systematics introduced by this choice of parameters are discussed in Section \ref{sec:5.3}.

For our purposes, namely computing half-light radii and fractional fluxes, we need to measure the Galaxy total flux from the images. Since its value depends on the image depth, we need to simulate the background noise. To do so, we first convert $I$ from $[\text{L}_{\sun}\text{ pc}^2]$ to $[\text{L}_{\sun}\text{ arcsec}^2]$ using the angular distance $D_\text{A}$. Then, we convert the surface brightness from $[\text{L}_{\sun}\text{ arcsec}^2]$ to $[\text{mag}\text{ arcsec}^2]$ using 
\begin{equation}
    \mu = -2.5\log I +5\log\left(\frac{D_\text{L}}{10\text{ pc}}\right) -2.5\log(1+z) + M_{\sun}
\end{equation}
where $\mu$ is the surface brightness in magnitude units, $D_\text{L}(z)$ is the luminosity distance, and $M_{\sun}$ is the absolute magnitude of the Sun, i.e. $4.51$ in the \textit{I}-band and $5.31$ in the \textit{B}-band \citep{2018ApJS..236...47W}. Finally, we chose a limiting magnitude $\mu_\text{lim}$ and set to $I=0$ and $\mu=\mu_\text{lim}$ all those pixels with $\mu>\mu_\text{lim}$. In our fiducial models we use $\mu_\text{lim}=25\,\text{mag}\text{ arcsec}^2$, motivated again by the typical values of the SGRB host observations. The edge-on and face-on images obtained with the fiducial setup are shown in Fig.~\ref{fig:1}.

The half-light radius $r_\text{e}$ is given as the semi-major axis of the isophote enclosing half of the total flux. We use elliptical isophotes, fitted using \texttt{isophote.Ellipse.fit\_image} from the \texttt{photutils} package \citep{larry_bradley_2023_7946442}. 

\section{Systemic and peculiar velocities}\label{sec:3}

Before analyzing the merger locations, it is worth taking a closer look at the BNS present-day velocities, as they bear some insight not only about the predicted merger locations but also about the BNS properties themselves. In the following Section we consider the 7 BNSs with measured proper motion, with the exception of B2127+11C which is likely bound to a GC, and thus has a proper motion which might be biased by the GC internal dynamics.

\subsection{Transverse velocity of the BNSs and their LSR}\label{sec:3.1}

\begin{figure}
	\centering
    \includegraphics[scale=1]{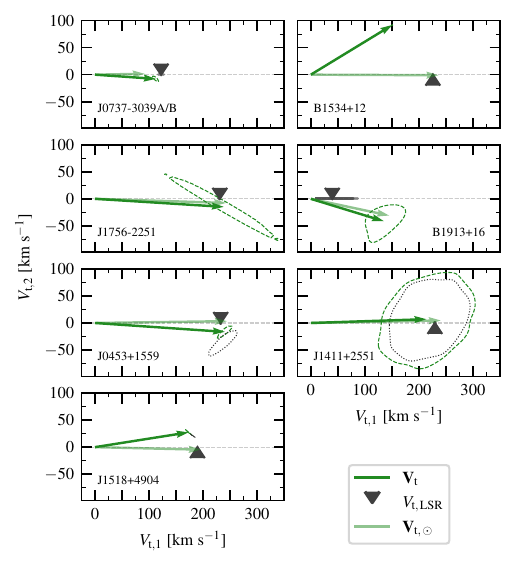}
    \caption{Transverse velocities of the BNSs in Fig.~\ref{fig:2}. The axes are arbitrarily oriented so that the LSR transverse velocity vectors ${\bf V}_\text{t,\,LSR}$ lie on the $x$-axis. Grey triangles indicate the LSR transverse velocities $V_\text{t,\,LSR}$, green arrows indicate the BNSs transverse velocities ${\bf V}_\text{t}$ , and light-green arrows indicate the transverse component of the Sun velocity ${\bf V}_\text{t,\,\sun}$. The dashed green lines enclose the $1\sigma$ regions of ${\bf V}_\text{t}$, while the grey shaded areas are the $1\sigma$ regions of $V_\text{t,\,LSR}$, both obtained from a Gaussian kernel density estimation of our MC simulations. These $1\sigma$ regions represent the uncertainties from the on-sky locations and proper motions, and the distance estimates. The dotted black lines of J0453+1559, J1411+2551, and J1518+4904, enclose the ${\bf V}_\text{t}$ $1\sigma$ regions when the DM distance is estimated from \citet{2002astro.ph..7156C} instead of \citet{2017ApJ...835...29Y}.
    }
    \label{fig:2b}
\end{figure}

Let us consider the BNS transverse velocities ${\bf V}_\text{t}$. When we compare them to the transverse velocities of their respective LSRs ${\bf V}_\text{t,\,LSR}$ (namely the circular velocity at their position), we find that the two have similar magnitudes and directions in all but two cases (see Fig.~\ref{fig:2b}). This suggests that the BNS systemic velocities are ($i$) not isotropically oriented in the Galactocentric frame, and ($ii$) not much different from the velocity of their LSR, or in other words, that they have small peculiar velocities with respect to their LSR. From Fig.~\ref{fig:2b}, we notice that this is due to the tangential component of the Sun's velocity ${\bf V}_\text{t,\,\sun}$ being aligned to ${\bf V}_\text{t,\,LSR}$, and dominating over the BNS proper motions. We also notice that ${\bf V}_\text{t,\,LSR}$ are almost completely in the tangential direction, since $V_\text{t,\,LSR}$ is $\sim200$ km s$^{-1}$ in all but two cases.

For the reason discussed in the previous paragraph, we conclude that the BNS systemic velocities inferred under the assumption of Galactocentric-isotropy (eq.~\ref{eq:1}) might be overestimated. Instead, those inferred under the assumption of LSR-isotropy (eq.~\ref{eq:2}) likely provide a lower limit, since similar ${\bf V}_\text{t}$ and ${\bf V}_{\text{t},\text{LSR}}$ would result in small radial velocities. This become clear when comparing the systemic and peculiar velocities $V$ and $V^\text{LSR}$ from the two assumptions. In the middle and bottom panels of Fig.~\ref{fig:2} we see that while we obtain similar $V$ from the two assumptions, the Galactocentric-isotropic $V^\text{LSR}$ all peak around $\sim200$ km s$^{-1}$ whereas the LSR-isotropic ones are distributed between $\sim10$ and $\sim200$ km s$^{-1}$. These latter estimates cover the same range of values as the BNS progenitors, namely NS-hosting high-mass X-ray binaries \citep[see Fig.~2 from][]{2022AA...665A..31F}, and that they get as low as the velocity dispersion in the thin disc \citep{2003AA...409..523R}.

The cumulative distributions of $V$ and $V^\text{LSR}$ for all the BNSs with measured proper motions (except the one in the GC, i.e. B2127+11C) are fitted with the lognormal distribution
\begin{equation}
    f(x,\alpha,\beta,\gamma)=\frac{1}{x\beta\gamma\sqrt{2\pi}}\exp\left[ -\frac{1}{2\gamma^2}\log^2\left( \frac{x-\alpha}{\beta} \right) \right]
\end{equation}
using \texttt{scipy.stats.lognorm.fit} \citep{2020SciPy-NMeth}. The fits are shown in the upper panels of Fig.~\ref{fig:2} and the fitted parameters are reported in Tab.~\ref{tab:3}. For the whole sample, we find $V\approx 285^{+190}_{-113}$ km s$^{-1}$ and $V^\text{LSR}\approx 182^{+209}_{-100}$ km s$^{-1}$ under Galactocentric-isotropy, and $V\approx 245^{+81}_{-56}$ km s$^{-1}$ and $V^\text{LSR}\approx 58^{+99}_{-36}$ km s$^{-1}$ under LSR-isotropy. This lower value is also consistent with that inferred by \cite{2016MNRAS.456.4089B}.

Despite the qualitative differences resulting from the two assumptions, we do not expect them to produce quantitative differences in the merger locations. Indeed, both assumptions result in systemic velocities that are close to the circular velocity \citep[$\sim230$ km s$^{-1}$, ][]{2012ApJ...759..131B,2017MNRAS.465...76M} and well below the escape velocity at the Sun location \citep[$\sim500-600$ km s$^{-1}$, ][]{2014AA...562A..91P,2018AA...616L...9M}. Therefore, they should lead to similar merger offsets. A quantitative analysis of their effect on the results is provided in Sec.~\ref{sec:5.1}.

\subsection{Implications for birth locations and velocities}\label{sec:3.2}

\begin{figure*}
	\centering
	\includegraphics[scale=1]{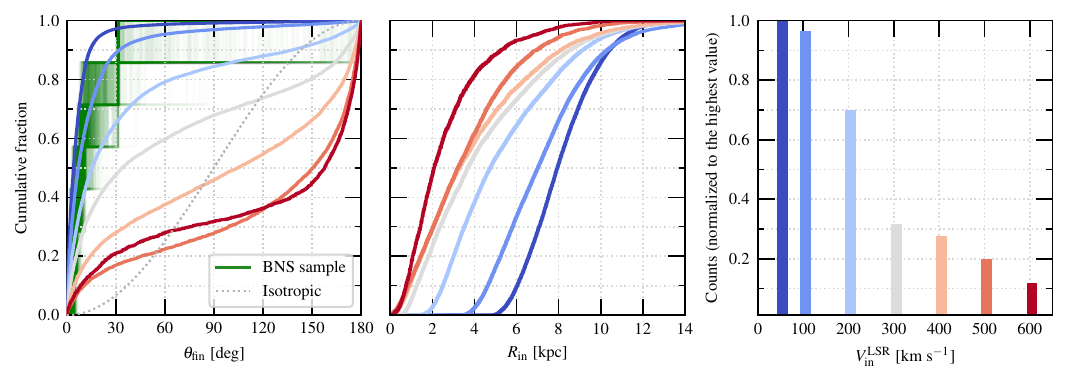}
    \caption{\textbf{Left.} Distribution of angles $\theta_\text{fin}$ between ${\bf V}_\text{t}$ and ${\bf V}_\text{t,\,LSR}$ for different kicks $V^\text{LSR}_\text{in}$. Green lines labelled "BNS sample" are realizations of the observed $\theta$ from Galactic BNSs, while all the other solid coloured lines are predictions for binaries in an exponential disc after an isotropic kick of given magnitude, as explained in Sec.~\ref{sec:3.2}. The color code for the kick magnitude $V^\text{LSR}_\text{in}$ is shown in the right plot. The grey dotted line is the distribution predicted for point masses with isotropic velocities, located at random positions within 1 kpc from the Sun. \textbf{Centre.} Initial radial positions $R_\text{in}$ of the point masses that end up within 1 kpc of the Sun for different kicks $V^\text{LSR}_\text{in}$. As we can see, high $V^\text{LSR}_\text{in}$ allow us to probe the inner Galactic regions, whereas low $V^\text{LSR}_\text{in}$ bias the sample to systems with $R_\text{in}\approx R_{\sun}$. \textbf{Right.} Counts of the point masses that end up within 1 kpc of the Sun for different kicks $V^\text{LSR}_\text{in}$, normalized to the highest value. As $V^\text{LSR}_\text{in}$ increases, the systems found around the Sun becomes less numerous.}
    \label{fig:5}
\end{figure*}

As mentioned in the previous Section, almost all the BNSs with proper motions have ${\bf V}_\text{t}$ closely aligned to ${\bf V}_\text{t,\,LSR}$, which hints that ${\bf V}$ is likely not isotropic. The deviation is evident when comparing the angles $\theta$ between ${\bf V}_\text{t}$ and ${\bf V}_\text{t,\,LSR}$ to the isotropic distribution, as shown in the left panel of Fig.~\ref{fig:5}. The ${\bf V}_\text{t}-{\bf V}_\text{t,\,LSR}$ alignment also implies that the BNSs might have peculiar velocities $V^\text{LSR}$ as low as those of their progenitors. This in turn suggests that our sample might be biased toward systems that received low kicks to the barycenter following the second supernova, and raises the question about how low the kicks should be in order to reproduce the observed $\theta$. 

To this end, we employ a toy model to check the range of suitable kicks. We simulate the trajectories of $10^{5}$ point masses with circular orbits in the Galactic disc after receiving a kick $V^\text{LSR}_\text{in}$. We test $7$ different kick magnitudes, namely $50$, $100$, $200$, $300$, $400$, $500$, and $600$ km s$^{-1}$, and for each of these values we seed the point masses in the plane at $Z=0$ using eq.~\ref{eq:5} as probability distribution. The masses are initialized with a velocity equal to the circular velocity plus a kick $V^\text{LSR}_\text{in}$ in a random direction. The trajectories are integrated for $5$ Gyr, and we record location and velocity at $300$ random times between $0$ and $5$ Gyr. Of all the locations we record, we select those that fall within $1$ kpc from the Sun since the BNS sample is biased to this region, and compute the angle $\theta_\text{fin}$ between ${\bf V}_\text{t}$ and ${\bf V}_\text{t,\,LSR}$. The respective starting radial positions $R_\text{in}$ are recorded as well.

The left panel of Fig.~\ref{fig:5} shows a comparison between the measured $\theta$ distribution and the predicted $\theta_\text{fin}$ distributions for each value of $V^\text{LSR}_\text{in}$. We find that lower kicks result in angles skewed toward small values (meaning ${\bf V}_\text{t}$ is mostly aligned to ${\bf V}_\text{t,\,LSR}$), while at higher kicks the angle distribution steepens at both low and high values (meaning ${\bf V}_\text{t}$ is either aligned or anti-aligned to ${\bf V}_\text{t,\,LSR}$). The measured $\theta$ distribution is well reproduced with kick magnitudes up to $100-200$ km s$^{-1}$, which are compatible with the values inferred from the observed SGRB offsets \citep[$\sim20-140$ km s$^{-1}$, ][]{2013ApJ...776...18F}. We find that strongly-kicked systems probe the inner regions of the Galaxy, while weakly-kicked systems probe only a region close to the Sun, as shown in the middle panel in Fig~\ref{fig:5}.

Since each realisation with different kicks in the disc start with the same number of binaries, the right panel of Fig.~\ref{fig:5} shows that the number of binaries with 1 kpc of us in the strong-kick scenarios is $<40$ per cent of the weak-kick scenarios. Therefore a modest population of strongly-kicked binaries would not be recovered in the BNS samples that we have to date. Taking into account that population synthesis predicts an anti-correlation between kicks and merger times \citep[e.g. Fig.~C1 in][]{2018MNRAS.481.4009V}, hence that strongly-kicked BNSs should be even rarer in the Milky Way given its age and low star formation rate, the ${\bf V}_\text{t}-{\bf V}_\text{t,\,LSR}$ alignment suggests that our sample probes only the BNS population born with small $V^\text{LSR}_\text{in}$ at around the same Galactocentric radius as the Sun. In other words, that our sample is neither representative of the BNSs born in the central regions, nor of the BNSs residing in the outer regions of the Milky Way.

\section{Merger locations}\label{sec:4}

In the following Sections we analyze the merger locations of the 5 Galactic BNSs with measured proper motions and $\tau_\text{gw}<14\text{ Gyr}$, over the fiducial Milky Way image in the \textit{I}-band. We refer to this setup as the fiducial model. The analysis employs three observables: the projected offsets $r_\text{h}$, the normalized offset $r_\text{n}$ and the fraction of light $f_\text{light}$. The first observable is the on-sky projection of the merger offset from the Galactic centre, the second is the projected offset expressed in units of $r_\text{e}$, while the third is the fraction of total light contained in pixels dimmer than the one at the transient location. The last two observables are commonly used in the analysis of transients locations on their hosts \citep[e.g.][]{2002AJ....123.1111B,2006Natur.441..463F}, including SGRBs \citep[e.g.][]{2014ARAA..52...43B}. 

We choose the viewing angles with which to project the 3D galaxy model onto a 2D image, for each merger, from an isotropic distribution. The possible values for $i$ and $\phi$ are distributed over a discrete grid with 10 values for $i$ and 20 for $\phi$. In particular, we use $i=\arccos(u)$ where the $u$ are 10 evenly-spaced values in $[0,1]$, and $\phi$ are 20 evenly-spaced values in $[0,\pi]$. 

The predictions are then compared to a sample of observed SGRBs, listed in Table~\ref{tab:4}. The sample is a subset of the SGRBs analyzed by \citet{2022ApJ...940...56F}, selected for having measured $r_\text{n}$. The $f_\text{light}$ values are collected from various works in the literature (see Table~\ref{tab:4} for the references).

\begin{table}
    \centering
    \caption{The SGRB sample used in this work. Columns list the host type, the redshift, the projected offset from the host centre in kpc $r_\text{h}$, the projected offset in units of the host's half-light radius $r_\text{n}$, and the fraction of light at the SGRB position $f_\text{light}$. Hosts are classified into star-forming (SF), transitioning (T), and quiescent (Q). Host types are taken from \citet{2022ApJ...940...57N}, while the remaining is from \citet{2022ApJ...940...56F}.
    }
    \label{tab:4}
    \begin{tabular}{lccccc}
    \hline\hline
    GRB & Type & $z$ & $r_\text{h}$ & $r_\text{n}$ & $f_\text{light}$ \\
    &&&[kpc]&[$r_\text{e}$]&\\
    \hline
    050509B & Q     & 0.23      & 55.19$\pm$12.43   & 2.59$\pm$0.58     & -    \\
    050709  & SF    & 0.16      & 3.76$\pm$0.056    & 2.00$\pm$0.03     & 0.09 $^\textrm{a}$ \\
    050724  & Q     & 0.25      & 2.74$\pm$0.08     & 0.67$\pm$0.02     & 0.33 $^\textrm{a}$ \\
    051210  & SF    & 2.58      & 29.08$\pm$16.34   & 5.65$\pm$3.17     & -    \\
    051221A & SF    & 0.55      & 2.08$\pm$0.19     & 0.89$\pm$0.083    & 0.65 $^\textrm{a}$ \\
    060121  & -     & -         & 0.97$\pm$0.37     & 0.18$\pm$0.069    & 0.41 $^\textrm{a}$ \\
    060313  & -     & -         & 2.60$\pm$0.55     & 1.39$\pm$0.3      & 0.00 $^\textrm{a}$  \\
    060614  & SF    & 0.13      & 0.70$\pm$0.79     & 0.86$\pm$0.97     & -    \\
    061006  & SF    & 0.46      & 1.39$\pm$0.29     & 0.37$\pm$0.077    & 0.63 $^\textrm{a}$ \\
    070429B & SF    & 0.90      & 6.00$\pm$13.44    & 1.17$\pm$2.62     & -    \\
    070707  & -     & -         & 3.25$\pm$0.24     & 1.11$\pm$0.083    & 0.00 $^\textrm{a}$ \\
    070714B & SF    & 0.92      & 12.33$\pm$0.87    & 5.17$\pm$0.37     & 0.00 $^\textrm{a}$ \\
    070724  & SF    & 0.46      & 5.52$\pm$0.18     & 1.49$\pm$0.048    & 0.23 $^\textrm{a}$ \\
    070809  & T     & 0.47      & 34.11$\pm$2.75    & 9.34$\pm$0.75     & -    \\
    071227  & SF    & 0.38      & 14.74$\pm$0.26    & 3.08$\pm$0.055    & 0.00 $^\textrm{a}$ \\
    080503  & -     & -         & 7.31$\pm$0.24     & 3.46$\pm$0.12     & -    \\
    090305  & -     & -         & 3.49$\pm$0.24     & 1.19$\pm$0.083    & 0.30 $^\textrm{a}$ \\
    090510  & SF    & 0.90      & 10.51$\pm$2.92    & 1.66$\pm$0.46     & 0.00 $^\textrm{a}$ \\
    090515  & Q     & 0.40      & 76.19$\pm$0.16    & 13.98$\pm$0.03    & 0.00 $^\textrm{a}$ \\
    091109  & -     & -         & 4.22$\pm$0.41     & 1.93$\pm$0.19     & -    \\
    100117A & SF    & 0.91      & 1.35$\pm$0.32     & 0.61$\pm$0.14     & 0.54 $^\textrm{a}$ \\
    130603B & SF    & 0.36      & 5.40$\pm$0.20     & 0.71$\pm$0.027    & 0.35 $^\textrm{a}$ \\
    130912A & -     & -         & 3.90$\pm$1.06     & 1.41$\pm$0.38     & -    \\
    131004A & -     & 0.72      & 0.80$\pm$0.22     & 0.25$\pm$0.068    & -    \\
    150101B & Q     & 0.13      & 7.36$\pm$0.072    & 0.78$\pm$0.0076   & 0.21 $^\textrm{b}$ \\
    150424A & -     & -         & 3.41$\pm$0.32     & 1.50$\pm$0.14     & -    \\
    160303A & SF    & 1.01      & 15.31$\pm$0.90    & 3.42$\pm$0.20     & -    \\
    160624A & SF    & 0.48      & 9.63$\pm$6.24     & 2.37$\pm$1.54     & -    \\
    160821B & SF    & 0.16      & 15.74$\pm$0.03    & 4.24$\pm$0.008    & -    \\
    170817A & Q     & 0.01      & 2.125$\pm$0.001   & 0.64$\pm$0.03     & 0.54 $^\textrm{c}$ \\
    200522A & SF    & 0.55      & 0.93$\pm$0.19     & 0.24$\pm$0.048    & 0.95 $^\textrm{d}$ \\
    211106A & -     & -         & 0.79$\pm$0.29     & 0.49$\pm$0.18     & -    \\
    211211A & SF    & 0.08      & 7.92$\pm$0.029    & 3.20$\pm$0.01     & -    \\
    \hline
    \end{tabular}
    \begin{flushleft}
    \textbf{References.} 
	   $^\textrm{a}$ \citet{2013ApJ...776...18F}, 
	   $^\textrm{b}$ \citet{Fong_2016}, 
	   $^\textrm{c}$ \citet{2017ApJ...848L..23F}, 
	   $^\textrm{d}$ \citet{2021ApJ...906..127F}
	\end{flushleft}
\end{table}

\subsection{Projected offsets}\label{sec:4.1}

\begin{figure*}
	\centering
    \includegraphics[scale=1]{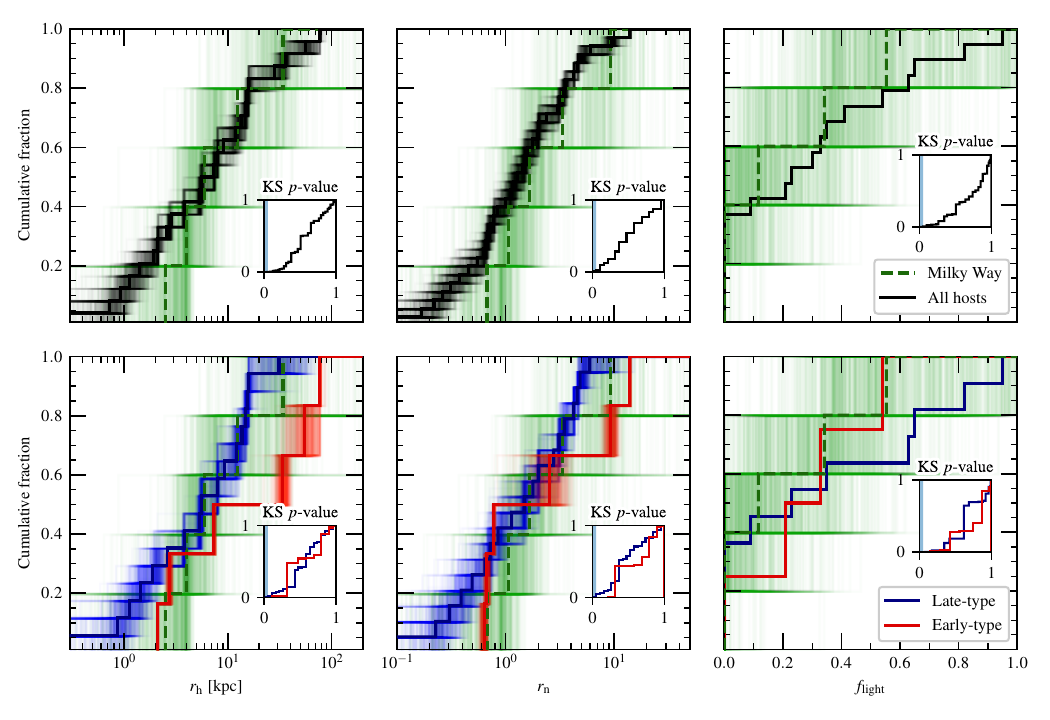}
    \caption{Projected offsets $r_\text{h}$ (left), normalized offsets $r_\text{n}$ (centre), and fraction of light $f_\text{light}$ (right) for BNS mergers and SGRBs. Solid thick lines represent the median distributions of SGRBs, either for the whole sample (upper panels) or divided by host type (lower panels). Dashed thick lines represent the median distributions of BNSs mergers from our fiducial model. Semi-transparent lines are the single realizations of the corresponding distributions. The insets show the distribution of the $p$-values from the KS test when comparing all the realizations one-to-one, with the shaded area indicating the $5\%$ region.}
    \label{fig:3}
\end{figure*}

We produce $10^4$ realizations of the $r_\text{h}$ distribution by picking 1 out of the $10^4$ merger locations for each of the 5 Galactic BNSs. The cumulative $r_\text{h}$ distributions are shown in the left panels of Fig.~\ref{fig:3}, together with those of the SGRBs. The latter are produced with a MC simulation, assuming the SGRB offsets in Tab.~\ref{tab:4} have Gaussian uncertainties with standard deviation equal to the observational uncertainties. We produce $10^4$ realizations of such distributions, so that we can compare them one-to-one to those of the Galactic BNSs. 

The Kolmogorov-Smirnov (KS) test cannot reject the null hypothesis at a significance level below $5$ per cent in virtually all realizations. In particular, the null hypothesis is rejected in $<0.3$ per cent of cases when comparing the BNS mergers to the whole SGRB sample, in $<0.5$ per cent of cases when comparing to SGRBs with late-types host (type Q and T in Tab.~\ref{tab:4}), and in $<0.1$ per cent when comparing to SGRBs with early-types host (type SF in Tab.~\ref{tab:4}). Note that the BNSs merger realizations are not independent, thus one should not expect to reject the null hypothesis at a $5$ per cent significance level in $\sim5$ per cent of cases, as per definition.

The median projected offset of the BNS mergers is $\langle r_\text{h}\rangle\approx 6$ kpc (see Fig.~\ref{fig:3}). This value is similar to the BNS present-day offsets, which due to observational bias is around the Galactocentric radius of the Sun, i.e. $R_{\sun}\approx8$ kpc (which is on average $20$ per cent smaller upon projection)\footnote{Consider a point on a sphere with cartesian coordinates $(x,y,z)=(\rho\cos\phi\sin i,\rho\sin\phi\sin i,\rho\cos i)$. The projected radius on the $xy$-plane is $R=\rho\sin i$. The expected value of $R$ assuming an isotropic random orientation is $\langle R \rangle=\frac{\int_0^{2\pi}{\rm d}\phi\int_0^\pi\rho\sin^2i\,{\rm d}i}{\int_0^{2\pi}{\rm d}\phi\int_0^\pi\sin i\,{\rm d}i}=\frac{\pi}{4}\rho\approx0.8\rho$.}, and it is a result of the systemic velocities being close to the circular velocity, and well below the escape velocity at the Sun location. By chance, these values are similar to the median projected offset of the SGRBs, which is $\langle r_\text{h}\rangle\approx 8$ kpc (see Fig.~\ref{fig:3}). When taken together with the small radial displacement in the BNS trajectories, this suggests that the similarity between the $r_\text{h}$ distributions of SGRBs and BNS mergers might be simply a coincidence, resulting from the BNS sample being biased toward systems located close to the Sun. Or in other words, if the Sun were located at a significantly larger (or smaller) Galactocentric radius, then we might find the same BNSs mergers peaking at a different $r_\text{h}$. Nevertheless, it is noteworthy that our models predict merger offsets that are comparable to the upper tail of the SGRB distribution despite this bias.

\subsection{Normalized offsets}\label{sec:4.2}

The cumulative distributions of $r_\text{n}$ for the BNS mergers are shown in the middle panels of Fig~\ref{fig:3}, together with those for the SGRBs. Similarly to the $r_\text{h}$, the KS test cannot reject the null hypothesis at a significance level below $5$ per cent in virtually all realizations. In particular, the null hypothesis is rejected in $<0.5$ per cent of cases when comparing the BNS mergers to the whole SGRB sample, in $<0.4$ per cent of cases when comparing to SGRBs with late-types host, and in $<0.1$ per cent when comparing to SGRBs with early-types host. Overall, we find that the projected offsets of the various samples agree better when normalized to $r_\text{e}$, as already noted by \cite{2022ApJ...940...57N}.

The inclusion of Milky Way's morphological properties in our analysis, e.g. $r_\text{e}$, raises the question about how representative our Galaxy is with respect to the SGRB host population. Although a thorough comparison is beyond the scope of this work, a brief discussion is already insightful. The Milky Way is a very bright yet relatively red spiral galaxy, which likely belongs to the green valley in the colour-magnitude diagram \citep{2011ApJ...736...84M,2015ApJ...809...96L,2020MNRAS.491.3672B}. Whereas its total stellar mass \citep[$M_{\star}\approx6\times10^{10}$ M$_{\sun}$,][]{2016ARAA..54..529B} is around the median value for transitioning/quiescent (T/Q) SGRB host and $1\sigma$ higher than star-forming (SF) hosts, its star formation rate \citep[$\dot{m}_{\star}\approx1.65$ M$_{\sun}$ yr$^{-1}$,][]{2016ARAA..54..529B} is $2\sigma$ higher than any T/Q host and below $\sim70$ per cent of the SF hosts \citep[see Fig.~5 from ][]{2022ApJ...940...57N}. Thus, we cannot conclusively compare the Milky Way with either Q/T or SF hosts alone, but we can conclude that our Galaxy is at least more massive than half of the Q/T hosts, and in the upper quartile of the whole host population. Despite this, it has been suggested that the scale length $R_\text{d}$ of the Milky Way disc is anomalously small \citep{1996ApJ...473..687M,2007ApJ...662..322H,2016ApJ...833..220L,2020MNRAS.491.3672B}. Based on the luminosity-velocity-radius (LVR) scaling relation, \cite{2016ApJ...833..220L} find that the Milky Way $R_\text{d}$ is half of the typical value of similar galaxies, which is $R_\text{d}\approx5$ kpc, and that our Galaxy lies farther from the LVR relation than $\sim90$ per cent of other spiral galaxies, in agreement with \cite{2007ApJ...662..322H}.

These remarks suggest that even if the Milky Way would be representative of the most massive SGRB hosts, the size of its stellar disc could play a role in shifting the $r_\text{n}$ distribution of BNS merger to lower values, supporting even more our claim that the agreement we find between BNS mergers and SGRBs is a cosmic coincidence. Nevertheless, we do not investigate the impact of a different $R_\text{d}$ since the stochastic spread in the predictions would still be dominant in the KS test.

\subsection{Fraction of light}\label{sec:4.3}

The cumulative distributions of $f_\text{light}$ for the BNS mergers are shown in the right-hand panels of Fig~\ref{fig:3}, together with those for the SGRBs. Similar to the previous cases, the KS test cannot reject the null hypothesis at a significance level below $5$ per cent in virtually all realizations. In particular, the null hypothesis is rejected in $<0.01$ per cent of cases when comparing the BNS mergers to either the whole SGRB sample, to SGRBs with late-types host, or to SGRBs with early-types host. We note that all the $f_\text{light}$ distribution is skewed to the left, meaning that both BNS mergers and SGRBs do not trace the stellar light, and that they are more likely found in the dimmer pixels. The remarks we made in the previous Section about the Milky Way $R_\text{d}$ have also implications for the $f_\text{light}$ distributions of BNS mergers, as a more compact disc would skew the distributions even more to the left. However, we do not test a different $R_\text{d}$ as the significant stochastic spread would still dominate in the KS test, as mentioned earlier.

\section{Systematics}\label{sec:5}

\subsection{DM distances}\label{sec:5.0}

To estimate the distance of J0453+1559, J1411+2551, and J1518+4904, we use their DM together with the model for the free-electron density in the Milky Way from \citet{2017ApJ...835...29Y}. Different electron-density models however can lead to different distance estimates, therefore we want to test the impact of our specific choice of model. To do this, we compare the results obtained with the model of \citet{2017ApJ...835...29Y} to those obtained with the widely-used model of \cite{2002astro.ph..7156C}.

First of all, we note that all three BNSs with DM distances have merger times greater than the Hubble time, therefore we do not use them for our fiducial models of $r_\text{h}$, $r_\text{n}$, and $f_\text{light}$. For this reason, the choice of a specific electron-density model do not impact our results on the BNS merger locations. Regarding the BNS velocities instead, we note that the two different models give similar results. The lognormal distributions fitted to both $V$ and $V^\text{LSR}$ have the same mean value and $1\sigma$ interval regardless of the electron-density model, as reported in Tab.~\ref{tab:3}. The angles $\theta$ between ${\bf V}_\text{t}$ and ${\bf V}_\text{t,\,LSR}$ are also not affected by a different electron-density model, as shown in Fig.~\ref{fig:2b}. We do not show a comparison between the $\theta$ distributions from the two electron-models as they overlap and would not be distinguishable. Therefore, we conclude that also our results involving the BNS systemic velocities are not affected by the choice of electron-density model.

\subsection{Isotropy assumptions}\label{sec:5.1}

The toy model discussed in Sec.~\ref{sec:3.2} has not only implications for how representative the BNS sample is of the whole Galactic population, but also for our estimates of the radial velocities $V_\text{r}$. The ${\bf V}_\text{t}-{\bf V}_\text{t,\,LSR}$ alignment disfavours the assumption of isotropy under which we obtain $V_\text{r}$ for our fiducial model, and also favours kicks with magnitudes less than or equal to the circular velocity, which means that the systemic velocities might still encode the Galactic rotation (although the velocity is not conserved during the orbit). 

To check the impact of the isotropy assumption, we perform the analysis in Sec.~\ref{sec:4.1}$-$\ref{sec:4.3} on models employing LSR-isotropic velocities instead of the Galactocentric-isotropic ones, while keeping all the other parameters unchanged. In this variation we compute $V_\text{r}$ by de-projecting the residuals $\|{\bf V}_\text{t}-{\bf V}_\text{t,\,LSR}\|$ on a random angle, thus simulating the extreme case in which the systemic velocities are the circular velocities at the BNSs locations plus some peculiar velocity. When comparing $r_\text{h}$, $r_\text{n}$, and $f_\text{light}$ of BNS mergers to those of SGRBs, the KS test gives the same results as for the fiducial model, namely it cannot reject the null hypothesis below a $5$ per cent significance level in virtually all the cases and for all three observables, even though the LSR-isotropic velocities decrease the higher values of $r_\text{h}$ and increase those of $f_\text{light}$. This result reflects the fact that even if Galactocentric- and LSR-isotropic velocities cover different ranges, they are still both close to the circular velocity and well below the escape velocity at the Sun's location, as discussed in Sec.~\ref{sec:3.1}. The distributions of observables predicted with by the two assumptions are compared in Fig.~\ref{fig:10}, where we see that the two deviates only for $r_\text{h}$ and $r_\text{n}$ at high values.

\subsection{Initial conditions}\label{sec:5.2}

\begin{figure}
	\centering
	\includegraphics{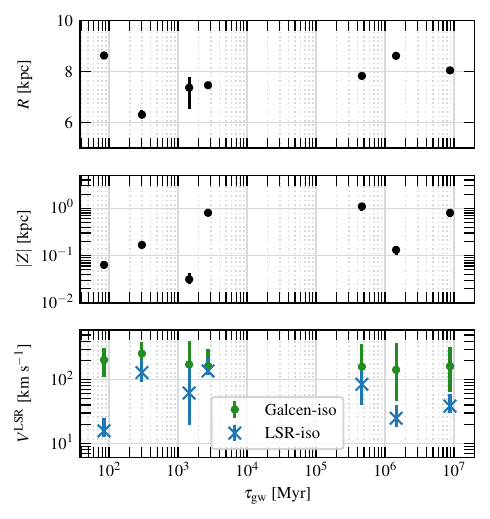}
    \caption{Merger times $\tau_\text{gw}$ against Galactocentric radii $R$, heights $Z$, and preculiar velocities $V^\text{LSR}$ for the BNSs in Fig.~\ref{fig:2}. The error bars extend from the 16th to the 84th percentiles, and are smaller than the marker when not visible.}
    \label{fig:9}
\end{figure}

Of the 8 confirmed BNSs with measured proper motion, only the 5 with $\tau_\text{gw}<14\text{ Gyr}$ have been used to predict the merger locations. We now employ the remaining 3 to understand how our results depends on the specific initial positions and velocities. To predict the merger locations of these 3, though, we cannot not use their true $\tau_\text{gw}$, since they are at least an order of magnitude greater than the Hubble time (see Table~\ref{tab:1}) and they might lead to unphysical offsets, i.e. in the case of unbound trajectories. Instead, we use one of the 5 BNSs $\tau_\text{gw}$ that are below 14 Gyr, motivated by the fact that the $\tau_\text{gw}$ do not show correlation with neither the BNSs positions nor their peculiar velocities, as shown in Fig.~\ref{fig:9}.

To do this, we repeat the analysis from Sec.~\ref{sec:4.1}$-$\ref{sec:4.3}, but this time we swap the initial conditions of each realization with the present-day position and velocity of a BNSs randomly drawn among the 8 with measured proper motions, while keeping the $\tau_\text{gw}$ unchanged. The KS test results remain the same of the fiducial model, namely the test cannot reject the null hypothesis below a $5$ per cent significance level in virtually all the cases, for all three observables and for both Galactocentric- and LSR-isotropic velocities. As shown in Fig.~\ref{fig:10}, this variation affects only the $f_\text{light}$ distribution skewing it more toward small values.

\subsection{Band and resolution of the synthetic image}\label{sec:5.3}

Lastly, we test our choice of parameters for the fiducial Milky way image. Since the vast majority of SGRBs observables we collect are obtained from \textit{HST} observations, we only test the effects of different redshifts together with different bands, without investigating the impact of PSF and pixel size. 

In the fiducial model, we modelled the Milky Way image using $z=0.5$ and the surface brightness in the $I$-band $\mu_I$. The choice of redshift is motivated by the median redshift of the SGRB hosts, while the choice of band is motivated by the SGRB hosts being observed mostly in red bands (e.g. \textit{HST} F814W ). This combination however is unphysical, since the observer-frame $I$-band correspond the rest-frame $V$-band for a galaxy at $z=0.5$. To test how it might impact our results, we repeat the analysis from Sec.~\ref{sec:4.1}$-$\ref{sec:4.3} with two different synthetic images, one with $z=0.2$ and $\mu_I$ and the other with $z=0.7$ and $\mu_B$. These two redshifts mark the 16th and 84th percentiles of the SGRB redshifts. For the higher redshift we use the $B$-band to mimic the bandshift, since light emitted in the $B$-band at $z=0.7$ would be observed approximately in the $I$-band. For the lower redshift we still use the $I$-band, which is a better approximation than the fiducial model since the observer-frame $I$-band correspond to the rest-frame $R$-band at $z=0.2$.

\begin{figure*}
	\centering
	\includegraphics{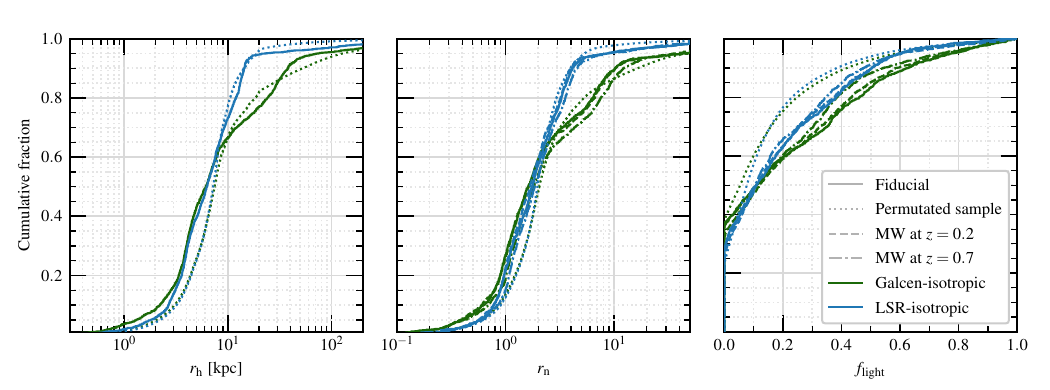}
    \caption{Distributions of observables predicted for the Galactic BNS mergers under several assumptions. Beside the predictions from the fiducial model, the other variations are meant to test different systematic effects that might arise from our methodology. Galactocentric- and LSR-isotropic distributions test the assumption of isotropy used to estimate the BNS radial velocities. Distributions from the permutated sample are obtained by permutating the initial conditions within the BNS that have measured proper motions but no constrain on the merger time, and are meant to test the dependency on the initial conditions within our sample. Distribution for the MW at $z=0.2$ and $z=0.7$ are obtained by changing band and angular resolution for the Milky Way image, to test the dependency on the image parameters.}
    \label{fig:10}
\end{figure*}

When comparing $r_\text{h}$, $r_\text{n}$, and $f_\text{light}$ of BNS mergers to those of SGRBs, the KS test gives the same results as for the fiducial model, namely it cannot reject the null hypothesis below a $5$ per cent significance level in virtually all the cases and for all three observables. This holds for both images, and for both Galactocentric- and LSR-isotropic velocities. Fig.~\ref{fig:10} shows that these variations do not produce significant differences in the predictions.

\section{Summary and conclusions}\label{sec:6} 

In this work we predicted the merger locations of the Galactic BNSs, and compared them to the locations of SGRBs on their hosts. We compared in particular the projected Galactocentric offsets $r_\text{h}$, the host-normalized offsets $r_\text{n}$, and the fraction of light $f_\text{light}$. 

Our fiducial model employs only 5 out of 15 confirmed BNSs, chosen for having measured proper motions and merger times $\tau_\text{gw}$ below the Hubble time. Their present-day Galactocentric positions and velocities are computed through a MC simulation that employs the on-sky positions and proper motions, distances estimated from the BNS dispersion measures, and radial velocities obtained by de-projecting the transverse velocity onto an isotropic orientation. The BNS trajectories are evolved in the Galactic potential starting from the present-day conditions up to $\tau_\text{gw}$. The merger locations are then analyzed on a synthetic image of the Milky Way, as if they were observed from a cosmological distance. The Galaxy model is composed of a bulge/bar plus a double-exponential disc, and the image is made for isotropic viewing angles in the $I$-band, assuming $z=0.5$, PSF FWHM of $0.1$ arcsec, pixel size of 0.05 arcsec px$^{-1}$, and limiting surface brightness of 25 mag arcsec$^{-2}$. 

When converting the present-day BNS proper motions into Galactocentric transverse velocities ${\bf V}_\text{t}$, we find that ${\bf V}_\text{t}$ are similar in magnitude and direction to the transverse velocity of each BNS Local Standard of Rest ${\bf V}_\text{t,LSR}$ in all but two cases. The similar directions suggest that BNSs have systemic velocities ${\bf V}$ which are not isotropically oriented in the Galactocentric frame, but that are rather aligned to the LSR velocity. The similar magnitudes suggest instead that BNSs have small peculiar velocities $V^\text{LSR}$ with respect to the LSR velocity. Using ${\bf V}_\text{t}$, we compute two different estimates for ${\bf V}$, one assuming ${\bf V}$ is isotropic, and the other assuming ${\bf V}^\text{LSR}$ is isotropic. We show that both systemic and peculiar velocities predicted for the observed Galactic BNSs fit a lognormal distribution.

Upon comparison with SGRBs, we find that our predicted BNS merger locations cover the same ranges of $r_\text{h}$, $r_\text{n}$, and $f_\text{light}$ as the SGRBs. We compare all the observables predicted for BNS mergers to those of the SGRBs with a KS test, which shows statistically non-significant differences in all cases. This could be attributed to the large spread in our predictions rather than the distributions being intrinsically similar. 

We test our results against a range of systematics that might be induced from our methodology. We find that the results from the fiducial model are robust against biases induced by the specific initial conditions of the trajectories, our estimates of the radial velocities, and our choice of parameters for the  Milky Way synthetic image. However, we find evidence that our BNS sample is not representative of the whole Galactic population, being biased toward systems that lies at around the same Galactocentric distance of the Sun, and have small peculiar velocities in their LSR. Thus, our sample is likely not representative neither of the BNSs born in the inner regions of the Galaxy, nor of those dwelling in the outer regions.

Although the connection between BNS mergers and SGRBs is supported by almost two decades of literature, we find that the agreement between the two shown by our analysis is non-trivial. The small peculiar velocities of the BNSs in our sample result in small radial displacements between the start and the end of their trajectories. That is to say, being all located at $R\approx 8$ kpc, they also merge at $R\approx 8$ kpc. Coincidentally, this is also the median offset of SGRBs. Furthermore, whereas the Milky Way is likely representative of the most massive SGRB hosts, its stellar disc might be more compact than similar spiral galaxies, which could result in lower $r_\text{n}$ and $f_\text{light}$. For this reason, we claim that the agreement we find between Galactic BNS mergers and SGRBs is likely a cosmic coincidence. Regardless, our results are noteworthy in the fact that we are still able to reproduce the highest values of $r_\text{h}$ and $r_\text{n}$, and the lowest values of $f_\text{light}$, for BNSs that start, travel, and merge close to the stellar disc. Also, our results suggest the need of further investigation into how representative the observed Galactic BNSs are compared to the whole Galactic population. A follow-up should analyze the observational biases characterizing the observed BNS sample, and could reveal new implications for the physical processes governing their systemic velocities.

\section*{Acknowledgements}

We thank the anonymous referee for the constructive comments. NG acknowledges studentship support from the Dutch Research Council (NWO) under the project number 680.92.18.02. AJL was supported by the European Research Council (ERC) under the European Union’s Horizon 2020 research and innovation programme (grant agreement No.~725246).
This work made use of Astropy\footnote{\url{http://www.astropy.org}}: a community-developed core Python package and an ecosystem of tools and resources for astronomy \citep{astropy:2013, astropy:2018, astropy:2022}.

\section*{Data Availability}

The data underlying this article are available at \url{https://gitlab.com/NicolaGsp/galbns_vs_sgrb}.



\bibliographystyle{mnras}
\bibliography{biblio} 







\bsp	
\label{lastpage}
\end{document}